\documentclass[]{jfm}

\usepackage{graphicx}
\usepackage{newtxtext}
\usepackage{newtxmath}
\usepackage{natbib}
\usepackage{hyperref}
\hypersetup{
    colorlinks = true,
    urlcolor   = blue,
    citecolor  = black,
}

\newcommand{\RomanNumeralCaps}[1]
\linenumbers

\usepackage{amsmath}
\usepackage{bm}
 \usepackage{comment}

\title{
Synchronisation in two-dimensional damped-driven Navier--Stokes turbulence: insights from data assimilation and Lyapunov analysis}

\author{Masanobu Inubushi\aff{1,2}
  \corresp{\email{inubushi@rs.tus.ac.jp}},
 \and Colm-cille P. Caulfield\aff{3,4}}

\affiliation{\aff{1}
Department of Applied Mathematics, Tokyo University of Science, Tokyo 162-8601, Japan
\aff{2}
Graduate School of Engineering Science, The University of Osaka, Osaka 560-8531, Japan
\aff{3}Institute for Energy and Environmental Flows, University of Cambridge, Cambridge CB3 0EZ, UK
\aff{4}Department of Applied Mathematics and Theoretical Physics, University of Cambridge, Cambridge CB3 0WA, UK}

\begin{document}
\maketitle

\begin{abstract}
In Navier--Stokes (NS) turbulence, large-scale turbulent flows inevitably determine small-scale flows. Previous studies using data assimilation with the three-dimensional NS equations indicate that employing observational data resolved down to a specific length scale, $\ell^{3D}_{\ast}$, enables the successful reconstruction of small-scale flows.
Such a length scale of `essential resolution of observation' for reconstruction $\ell^{3D}_{\ast}$ is close to the dissipation scale in three-dimensional NS turbulence.
Here we study the equivalent length scale in  {\it two}-dimensional NS turbulence, $\ell^{2D}_{\ast}$,
and compare with the three-dimensional case.
Our numerical studies using data assimilation and conditional Lyapunov exponents reveal that,
for Kolmogorov flows with Ekman drag,
the length scale $\ell^{2D}_{\ast}$ is actually close to the forcing scale, substantially larger than the dissipation scale.
Furthermore, we discuss the origin of the significant relative difference between the length scales, $\ell^{2D}_{\ast}$ and $\ell^{3D}_{\ast}$,
based on inter-scale interactions, `cascades' and orbital instabilities in turbulence dynamics.
\end{abstract}


\section{Introduction}
\begin{figure}
  \centerline{\includegraphics[width=0.7\linewidth]{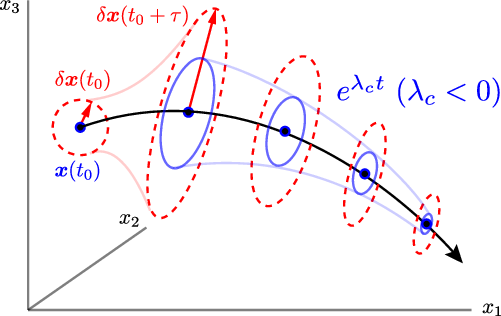}}
  \caption{Schematic of orbital instability and data assimilation in a chaotic dynamical system.
Orbital instability expands uncertainty (dashed ellipses), whereas data assimilation contracts uncertainty (solid ellipses) along the trajectory.
The figure illustrates a case in which data assimilation succeeds, with the uncertainty contracting exponentially along the orbit,
characterised by the negative conditional Lyapunov exponent $\lambda_c~(<0)$, as $e^{\lambda_c t}$.}
\label{fig1}
\end{figure}

Sensitive dependence on initial conditions is one of the crucial properties of turbulence dynamics.
Considering turbulence as a chaotic dynamical system, we can characterise it using the (maximum) Lyapunov exponent, which measures the speed of exponential growth of uncertainty in a state space.
Let us consider a dynamical system defined by differential equations, $\dot{x}={F}(x)$, with some initial condition, ${x}(t_{0})={x}_{0} \in \mathbb{R}^{N}$.
Precise data on the initial conditions ${x}_{0}$ are not available in practice due to the limitation of measurement, and so it is natural to introduce an initial uncertainty around $x_{0}$.
Schematically, the uncertainty is shown as a red ball in figure~{\ref{fig1}}.

In a chaotic dynamical system, uncertainty grows exponentially fast in the most unstable direction, $\delta x (t_{0}+\Delta \tau)$, depicted by a red
arrow and ellipsoid in figure {\ref{fig1}},
and $\| \delta x (t_{0}+\Delta \tau) \| \propto \| \delta x (t_{0}) \| e^{\lambda_1 \Delta \tau}$
where $\lambda_1~(>0)$ is the (maximum) Lyapunov exponent.
When we consider the above differential equations as the Navier--Stokes (NS) equations,
the positive Lyapunov exponent $\lambda_1$ indicates the sensitive dependence on initial conditions of turbulence,
which makes prediction difficult.
Specifically, it is conjectured that, for three-dimensional turbulence, the positive Lyapunov exponent is the inverse of the Kolmogorov time $(\nu/\epsilon)^{1/2}$, where $\nu$ is the kinematic viscosity and $\epsilon$ is the kinetic energy dissipation rate \citep{Ruelle}; thus, uncertainty grows with the fastest time scale in turbulence.

What if we could observe the state? Although complete observation of the state would eliminate uncertainty, it is unrealistic in practice. Instead, it is natural to assume that we can observe large-scale structures of turbulence. In other words, we can make {\it partial} or {\it incomplete} observation, since we can only observe at a relatively low resolution, where 
 the key control parameter is the resolution of our observations. 
The wavenumber corresponding to the resolution is denoted by $k_{a}$, and
we assume that data on a low-pass filtered field with $k<k_{a}$ are available.

If the resolution is too low, i.e., $k_{a}$ is too small, the chaotic dynamics would still expand the uncertainty along the orbit.
However, if we can obtain observational data with sufficiently high resolution, i.e., $k_{a}$ is sufficiently large,
the introduction of observational data overcomes chaotic dynamics, and reduces the uncertainty in the state, which is represented by the blue ellipsoid in figure {\ref{fig1}}.
This is an interpretation of {\it data assimilation} (DA) from the viewpoint of dynamical systems theory.
The outcome of DA, whether successful or not, is determined by the competition between uncertainty {\it expansion} by chaotic dynamics and uncertainty {\it reduction} through assimilation using observational data.
Interestingly, for three-dimensional turbulence in a periodic box, if $k_{a}$
exceeds a critical value, $k_{a} \ge 
k_{a}^{\ast(3D)}:=0.2/\eta$ where $\eta$ is the Kolmogorov length scale, the assimilation process progressively reduces and eventually eliminates the uncertainty.
In other words, the small-scale structure associated with $k \ge k_{a}$
converges asymptotically to the true unobserved structure under the DA process.
This critical value is less dependent on the details of external forcing and DA algorithms.
Since Yoshida, Yamaguchi and Kaneda (2005) first discovered this property  through direct numerical simulations in a periodic box, the DA approach
has attracted increasing attention and has been extended to various three-dimensional turbulent systems, including wall turbulence (see \cite{Zaki} and references therein).
Moreover, the synchronisation property of turbulence plays a key role in applications. For example, the above critical wavenumber, $k_{a}^{\ast(3D)}$,
is associated with the stability transition of machine learning-based turbulence models \citep{MIG}.

The competition inherent in the DA process can be mathematically formulated as a bifurcation problem, specifically referred to as the `blowout bifurcation' in the context of chaos synchronisation \citep{ISKG}.
Within this framework, the rate of uncertainty growth or reduction in the DA process can be quantified using the {\it conditional} Lyapunov exponent, denoted by $\lambda_{c}$ (see figure {\ref{fig1}}).
In particular, the critical wavenumber, $k_{a}^{\ast}$ can be identified by the sign change of the conditional Lyapunov exponent,
which also plays a central role in the present study in analysing the synchronisation.

Turbulence dynamics strongly depends on the spatial dimension. 
In this paper, we demonstrate that a remarkable distinction between two- and three-dimensional turbulence emerges in their synchronisation properties.
Regarding  DA in the two-dimensional NS equations,
a seminal paper by \cite{Titi} rigorously established a sufficient condition for  successful DA, based on the concept of determining modes~\citep{FMRT}, and this led to new DA schemes with more general observables.
Despite the critical importance of their series of results, they appear to be under-recognised within the turbulence physics community.
Here we revisit this problem with the novel tool of the conditional Lyapunov exponent. In particular, we identify the critical wavenumber in two-dimensional turbulence,
$k_{a}^{\ast(2D)}$, compare it with the equivalent critical wavenumber in three-dimensional turbulence, $k_{a}^{\ast(3D)}$, and discuss the underlying key physical differences in the dominant flow dynamics.
In \S 2, we present the mathematical formulation of the problem and describe the numerical methods.
The principal contributions of this study are twofold:
the numerical findings for Kolmogorov flow with Ekman drag, presented in \S3, and the physical interpretation of their dimensional dependence, proposed in \S4.
Finally, in \S 5, we provide concluding remarks and discuss directions for future work.

\section{Formulation and methods}
\label{sec:headings}

Here, we present the mathematical formulation of  DA for the NS equations, following \cite{Titi}.
In addition to single-run DA experiments, we introduce the conditional Lyapunov exponent as a long-time averaged quantity of the equations, based on \cite{ISKG}.

\subsection {Continuous Data Assimilation of the Navier--Stokes equations}

Two-dimensional turbulence is significant not only in mathematical physics 
but also as a basis for understanding (for example) atmospheric and oceanic flows.
In particular, we focus on Kolmogorov flows as an extremely simplified model.
However, two-dimensional NS flows driven by a forcing term consisting of a single eigenmode of the Laplacian,
including Kolmogorov flows, cannot sustain turbulence with a broad energy spectrum (Constantin, Foias \& Manley 1994).
Here, we introduce  Ekman drag, which is a natural physical mechanism to prevent energy accumulation in low-wavenumber regions (Boffetta \& Ecke 2012).
Kolmogorov flow with Ekman drag has also been widely used as a benchmark test in machine learning applications (for example, \cite{SCVM}).
Therefore, a precise understanding of the dynamics of this system is crucial across various fields of fluid mechanics.

Kolmogorov flow with Ekman drag is described by the NS equations with an incompressible velocity field,
\begin{align}
\frac{\partial {\bm u}}{\partial t} + {\bm u} \cdot \nabla {\bm u} = - \nabla \pi + \nu \Delta {\bm u} - \alpha {\bm u} + {\bm f},~~~\nabla \cdot {\bm u}=0,\label{NSE}
\end{align}
where ${\bm u}({ \bm x},t)$ and $\pi({\bm x},t )$ is the velocity and pressure field on a two-dimensional torus $\mathbb{T}^{2}=[0, 2 \pi]^{2}$, i.e., ${\bm u}: \mathbb{T}^{2} \times [0, \infty] \to \mathbb{R}^{2}$
and $\pi: \mathbb{T}^{2} \times [0, \infty] \to \mathbb{R}$.
The external forcing is of so-called Kolmogorov type, ${\bm f}=\sin (k_{f} y) {\bm e}_{x}$,
$\nu$ and $\alpha$ are the kinematic viscosity and the friction coefficient of the Ekman drag, respectively.
As in \cite{Titi}, the spatial mean of ${\bm u}$ over $\mathbb{T}^2$, i.e., the mean flow, is set to zero.

We introduce the projections $P_{k_{a}}$ and $Q_{k_{a}}$ in the Fourier representation as a mathematical model of incomplete observation, which is defined by
\begin{align}
P_{k_{a}} {\bm u}= \sum_{| {\bm k} | < k_{a}} \hat{\bm u}_{\bm k} e^{i {\bm k} \cdot {\bm x}}~~~{\text {and}}~~~
Q_{k_{a}}= I - P_{k_{a}},
\label{def_ka}
\end{align}
where $\hat{\bm u}_{\bm k}$ is the Fourier coefficient of the velocity field ${\bm u}$ for the wavenumber ${\bm k} \in \mathbb{Z}^{2}$.
This defines the observation of large-scale structures discussed in \S 1.
We assume that
the low-pass filtered field is observable, i.e., ${\bm p}:=P_{k_{a}} {\bm u}$,
while the unobserved small-scale velocity field corresponds to ${\bm q}:=Q_{k_{a}} {\bm u}$.
Note that the key parameter $k_{a}$ controls the resolution of the observation as it defines the orthogonal decomposition of the velocity fields into the large-scale and small-scale structures, $\bm u = {\bm p} + {\bm q}$.
For notational simplicity, we hereafter denote the equations (\ref{NSE}) by
\begin{align}
\frac{\partial {\bm u}}{ \partial t} = {\bm N}({\bm p}, {\bm q}), \label{simpleNSE}
\end{align}
and, by using the decomposition, the evolution equations for the large-scale structures field $\bm p$ and the small-scale structures field $\bm q$ are
\begin{align}
\frac{\partial {\bm p}}{ \partial t} = P_{k_{a}} {\bm N}({\bm p}, {\bm q})=: {\bm F}({\bm p}, {\bm q})~~~{\text{and}}~~~
\frac{\partial {\bm q}}{ \partial t} = Q_{k_{a}} {\bm N}({\bm p}, {\bm q})=: {\bm G}({\bm p}, {\bm q}).
\label{pqeqn}
\end{align}

The goal is to infer the unobserved small-scale field $\{ {\bm q}(t) \}_{t \ge 0}$ from the observational data of $\{ {\bm p}(t) \}_{t \ge 0}$.
To this end, we employ the method of continuous data assimilation (CDA).
We introduce an approximation velocity field, $\tilde{\bm u}$,
which is decomposed as $\tilde{\bm u}=\tilde{\bm p}+\tilde{\bm q}$ in the same way as ${\bm u}$.
In the CDA process, using the observational data, we set $\tilde{\bm p} (t)\equiv {\bm p} (t)$ for $\forall t \ge 0$,
a process called {\it direct insertion}, and then solve the relevant equations for the approximate small-scale structures field within the CDA framework:
\begin{align}
\frac{\partial \tilde{\bm q}}{ \partial t} =  {\bm G}({\bm p}, \tilde{\bm q}) \label{qtilde}
\end{align}
with some initial values $\tilde{\bm q} (0) = \tilde{\bm q}_{0}$.
We expect that if $k_{a}$ is sufficiently large, using the CDA process, that the approximation velocity field will converge to the actual velocity field, or equivalently 
\begin{align}
\tilde{\bm q}(t) \to {\bm q}(t)~~~(t \to + \infty) \label{conv}
\end{align}
from an {\it arbitrary} initial value, as confirmed in  three-dimensional turbulence \citep{Zaki}.
In such a case, we say that the small-scale flow ${\bm q}(t)$ is {\it synchronised} or {\it slaved} to the large-scale flow ${\bm p}(t)$, since the sequential data of $\{ {\bm p} (t)\}_{t\ge0}$
uniquely determine the asymptotic state of $\{ {\bm q} (t)\}_{t\ge0}$ after the transient period.

\subsection {Conditional Lyapunov exponent}\label{subsecCLE}

The notion of uncertainty is formulated geometrically by a perturbation vector to the velocity field, i.e., ${\bm u} + \delta {\bm u}$, where $\delta {\bm u}$ represents an error in measurement.
Let us consider the decomposition of the perturbation as $\delta{\bm u}= \delta {\bm p} + \delta {\bm q}$, as before.
In the CDA framework, we assume that there is no uncertainty in large-scale flows, as it is directly observable.
This assumption is mathematically equivalent to the condition $\delta {\bm p} (t)\equiv {\bm 0}$ for $\forall t \ge 0$.
Therefore, the uncertainty exists only in small-scale flows, i.e., ${\bm q} + \delta {\bm q}$.
In a case where the  CDA process has been successful, formulated by the condition (\ref{conv}),
the synchronised state, i.e., the true small-scale structure field $\bm q$, is expected to be asymptotically stable.
Hence, if we write $\tilde{\bm q}= {\bm q} + \delta {\bm q}$, 
the condition (\ref{conv}) is equivalent to $\delta {\bm q}(t) \to {\bm 0}~~(t \to + \infty)$ for an arbitrary initial perturbation vector field, $\delta {\bm q}(0)= \delta {\bm q}_{0}$.

To perform linear stability analysis, we further assume that the perturbation is infinitesimally small.
From the relation, $\tilde{\bm q}= {\bm q} + \delta {\bm q}$, with the evolution equations of ${\bm q}$ in (\ref{pqeqn}),
the linearisation of the equations (\ref{qtilde}) results in
\begin{align}
\frac{\partial \delta {\bm q}}{ \partial t} = D_{\bm q} {\bm G}({\bm p}, {\bm q}) \delta {\bm q},
\label{ddqdt}
\end{align}
where $D_{\bm q} {\bm G}$ is the Fr\'echet derivative
and $({\bm p}(t), {\bm q}(t))$ is the solution of the Navier--Stokes equations 
(\ref{simpleNSE})
on the turbulence attractor.
If the norm of the perturbation decays, $ \| \delta {\bm q}(t)  \| \to 0~(t \to + \infty)$,
the uncertainty vanishes, which means that the CDA has been successful.
Exponential convergence is characterised by the {\it conditional} Lyapunov exponent (CLE), 
\begin{align}
\lambda_{c}:= \lim_{T \to + \infty} \frac{1}{T} \ln \frac{ \| \delta {\bm q}(T) \|}{ \|  \delta {\bm q}(0) \|}
\label{defCLE}
\end{align}
if the limit exists.
It is an ergodic quantity that characterises the turbulence attractor of the NS equations.
In particular, it does not depend on any specific DA algorithm or the initial conditions.
A negative CLE indicates convergence of the CDA process independently of the initial point on the attractor, in contrast to the outcome of a single DA experiment, which may depend on the initial condition.

From its definition, it is clear that the CLE, $\lambda_c$, depends on the assimilation wavenumber $k_a$, which specifies the decomposition of ${\bm p}$ and ${\bm q}$. To make this dependence explicit, we denote it by $\lambda_c(k_a)$.
The CLE for the case of $k_a=1$ coincides with the maximum Lyapunov exponent,
i.e., $\lambda_c(k_a=1)=\lambda_1$, as discussed in \cite{ISKG}.
This is because, for $k_a=1$, we have ${\bm p} = P_{k_a=1}{\bm u} = \hat{\bm u}_{\bm 0}$ from the definition (\ref{def_ka}).
Since we consider flows with zero mean, the Fourier coefficient of the zero mode vanishes; hence, $\hat{\bm u}_{\bm 0} = {\bm 0}$.
Consequently, the case of $k_a=1$ corresponds to the situation without observational data, where $\bm p={\bm 0}$ and ${\bm u}={\bm q}$.
In this case, no condition is imposed on the perturbation vector, and thus the CLE reduces to the maximum Lyapunov exponent, $\lambda_c(k_a=1)=\lambda_1$.

\subsection{Numerical methods}

The direct numerical simulations are performed using a Fourier spectral method with $128 \times 128$ grid points, dealiasing implemented via the $3/2$ rule. Time integration is carried out using a fourth-order Runge--Kutta scheme.
In the next section, we first present the results for the case with forcing wavenumber $k_f = 4$, viscosity $\nu = 1.0 \times 10^{-3}$, and linear drag coefficient $\alpha = 1.0 \times 10^{-1}$, as the reference case, followed by an investigation of the dependence of the results on these parameters.

\begin{figure}
  \centerline{\includegraphics[width=\linewidth]{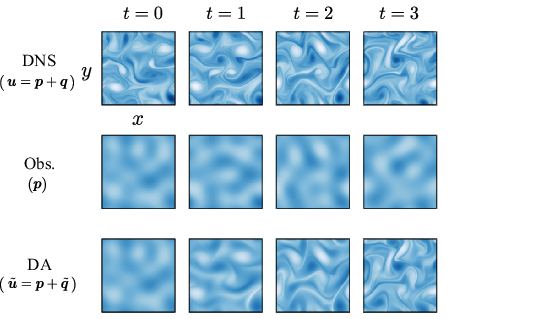}}
  \caption{
  Data assimilation experiment for two-dimensional Navier--Stokes turbulence.
  Snapshots of vorticity fields at times $t = 0, 1, 2$, and $3$, from top to bottom: the reference DNS field $\bm{u}(t) = \bm{p}(t) + \bm{q}(t)$; the observational data $\bm{p}(t)$ obtained by low-pass filtering the DNS field with $k_a = 4$; and the field obtained through data assimilation (CDA) $\tilde{\bm{u}}(t) = \bm{p}(t) + \tilde{\bm{q}}(t)$. A corresponding video is available as supplementary material.}
\label{fig2}
\end{figure}

\section{Results}\label{Results}

Figure~{\ref{fig2}} presents the results of simulations implementing the CDA process, showing snapshots of the vorticity field, $\omega= \partial_x v - \partial_y u$.
The top row, labelled `DNS', shows the results from  direct numerical simulations (DNS) of equations (\ref{NSE}),
that is, ${\bm u}(t)=(u(t),~v(t))^T={\bm p}(t) + {\bm q}(t)$ for $0 \le t \le 3$,
taken after the flow has reached a statistically steady state.
We also present the results of a simulation implementing the CDA process with $k_a=4$, initiated at $t=0$.
The middle row of figure~{\ref{fig2}}, labelled `Obs.', represents the vorticity field of the observation, ${\bm p}(t)$, 
obtained by applying a low-pass filter to the DNS data.

The bottom row of figure~{\ref{fig2}}, labelled `DA', represents the vorticity field constructed from the approximate velocity fields via $\tilde{\omega} = \partial_x \tilde{v} - \partial_y \tilde{u}$ where $\tilde{\bm u}(t)={\bm p}(t) + \tilde{\bm q}(t)$ are the results of the simulations implementing the CDA process.
The approximate small-scale structure field $\tilde{\bm q}(t)$ is computed by numerically solving equations (\ref{qtilde}) with the initial condition of $\tilde{\bm q}(0)=\tilde{\bm q}_0 = 0$,
with ${\bm p}(t)$ determined from the DNS results.
At the onset of the CDA process, $t=0$, only a single snapshot of the observation is available, and so $\tilde{\bm u} (0)= {\bm p}(0)$.
However, the simulation implementing the CDA process, using successive observational data ${\bm p}(t)$, gradually reconstructs the unobserved field ${\bm q}(t)$,
suggesting that $\tilde{\bm q}(t) \to {\bm q}(t)~(t \to +\infty)$.
In particular, at $t=3$,
the filamentary structures in the reconstructed vorticity field corresponding to $\tilde{\boldsymbol{u}}(t)$ closely resemble those in the DNS field, ${\bm u}(t)$.

\begin{figure}
  \centerline{\includegraphics[width=0.8\linewidth]{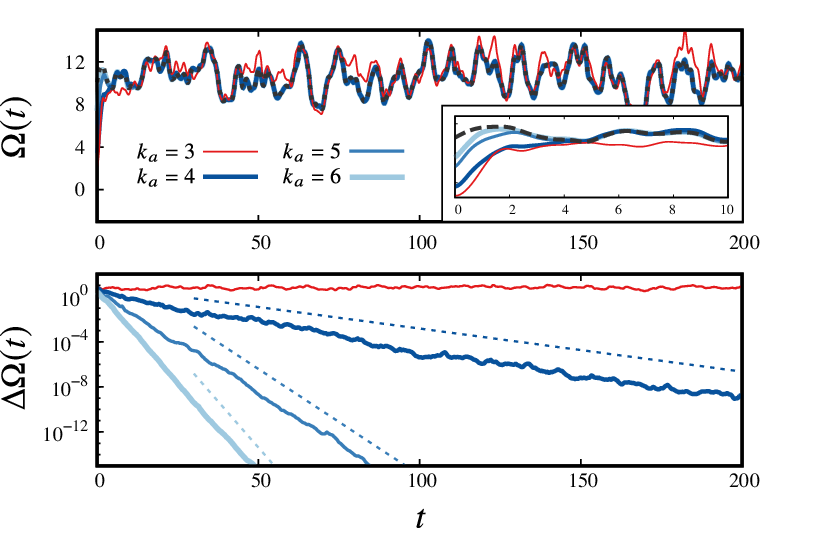}}
  \caption{Time series of enstrophy $\Omega(t)$ (top), with a close-up view for $0 \le t \le 10$ (inset), and the enstrophy-norm error $\Delta\Omega(t)$ as defined in (\ref{eq:errordef}) (bottom). The dashed lines show the convergence rates given by the conditional Lyapunov exponents $\lambda_c(k_a)$.
}
\label{enstrophy}
\end{figure}

\begin{figure}
  \centerline{\includegraphics[width=0.8\linewidth]{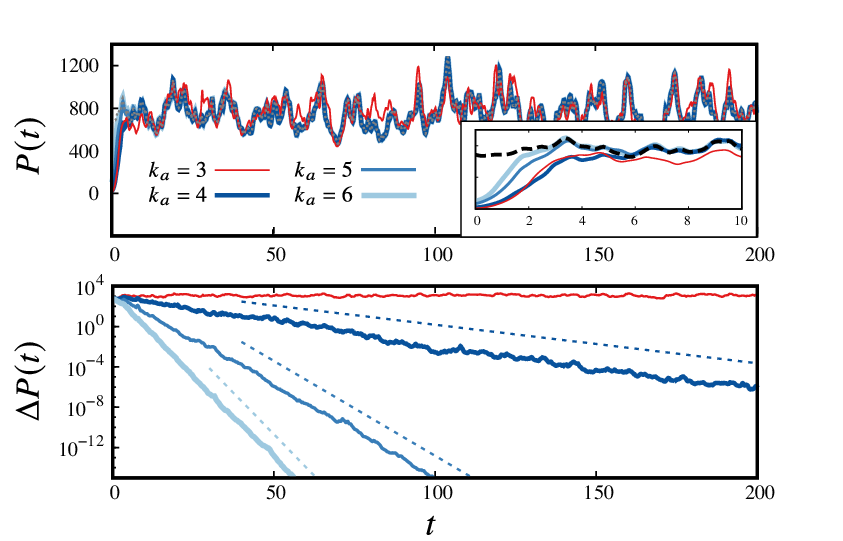}}
  \caption{Same as figure~{\ref{enstrophy}}, with the vertical axis showing the palinstrophy, $P(t)$, and the palinstrophy-norm error, $\Delta P(t)$.}
\label{palinstrophy}
\end{figure}

Figure~{\ref{enstrophy}} shows time series of enstrophy, $\Omega(t) = \frac{1}{2} \| \omega(t) \|_2^2$.
The dashed line corresponds to the DNS result, while the solid lines represent the enstrophy of the CDA reconstructions (i.e., $\frac{1}{2} \| \tilde{\omega}(t) \|_2^2$).
The line thickness increases with the wavenumber of $k_a$, showing the results for $k_a = 3, 4, 5, 6$ (with $k_a = 4$ corresponding to the case shown in figure~{\ref{fig2}}).
When the resolution is low, particularly for $k_a = 3$, the enstrophy shows a clear deviation from the DNS result; however, for $k_a \geq 4$, the CDA enstrophy converges towards the true value.
To examine this convergence in detail, we define the approximation error as 
\begin{equation}
\Delta\Omega(t) = \frac{1}{2} \| \omega(t) - \tilde{\omega}(t) \|_2^2, \label{eq:errordef}
\end{equation}
and plot this error in the lower panel of figure~{\ref{enstrophy}} using the same line thickness and colour as in the upper panel (the case $k_a = 4$ again corresponds to figure~{\ref{fig2}}).
For $k_a \geq 4$, the approximation error decays exponentially so $\Delta\Omega(t) \to 0 \quad (t \to \infty)$.

In order to characterise these simulation results in terms of the ergodic property of the NS equations,
we compute the conditional Lyapunov exponent $\lambda_c$ using variational equations (\ref{ddqdt}) for each $k_a$.
The dashed lines in the bottom panel of figure~{\ref{enstrophy}} show the exponential decay $\propto e^{2\lambda_c(k_a)t}$. The line slopes are in excellent agreement with the average decay rate of the approximation error $\Delta\Omega(t)$.
In particular, we obtain $\lambda_c(k_a) <0$ for $k_a \ge 4$, while $\lambda_c (k_a)>0$ for $k_a \le 3$.
Although specific properties, such as the convergence behaviour of the simulations implementing the CDA process  depend on the initial state ${\bm u}(0)$ on the turbulence attractor
and the initial guess of $\tilde{\bm q}(0)$, the conditional Lyapunov exponent does not.
Thanks to this property, we can conclude that, for Kolmogorov flow with $k_f=4$ studied here,
the critical wavenumber $k_{a}^{\ast(2D)}$ for synchronisation satisfies $3 < k_{a}^{\ast(2D)} < 4$.
In particular, the critical wavenumber is close to the forcing wavenumber; that is, 
$k_{a}^{\ast(2D)} \simeq {O}(k_f)$.

Figure~{\ref{palinstrophy}}
shows the same time series as in figure~{\ref{enstrophy}}, but in terms of the palinstrophy,
i.e., the upper panel shows $P(t)= \frac{1}{2} \| \nabla \omega(t) \|_2^2$,
and the lower panel shows $\Delta P(t) = \frac{1}{2} \| \nabla (\omega(t) - \tilde{\omega}(t)) \|_2^2$.
While the palinstrophy is more suitable for characterising  small scales in 2D turbulence,
the results are essentially the same as those obtained in terms of the enstrophy,
since all norms are equivalent in the finite-dimensional system, including the Galerkin-truncated system studied here.

\begin{figure}
  \centerline{\includegraphics[width=0.9\linewidth]{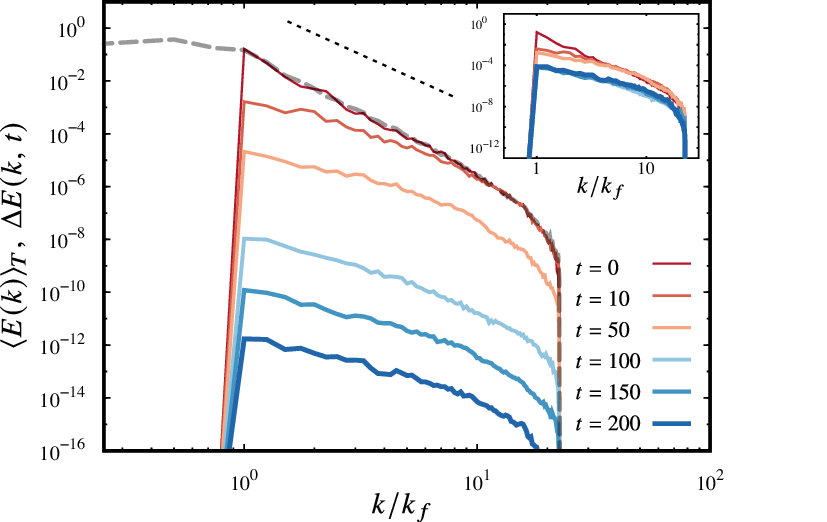}}
  \caption{
Synchronisation process examined over scales.
The solid lines show the time evolution of the energy spectrum of the difference of the fields $\Delta E(k,\,t)$
obtained through the CDA process for the case $k_a = 4$.
From thickest to thinnest, the curves correspond to times $ t = 0,\, 10,\, 50,\, 100,\, 150,\, \text{and}\ 200$.
The dashed line shows the result of the DNS, corresponding to the long-time-averaged energy spectrum $\langle E(k) \rangle_T$.
The inset shows the scaled energy spectra, $\Delta E(k,\,t)e^{2 \lambda_c t}$, plotted with the same colours.
}
\label{fig3}
\end{figure}

Figure~{\ref{fig3}} shows the energy spectrum illustrating the synchronisation process across scales.
The grey dashed line represents the time-averaged energy spectrum $\langle E(k) \rangle_T$ computed from the DNS data.
A power law region can be identified,
whose slope is steeper than the slope of the so-called Batchelor--Kraichnan--Leith scaling $E(k) \propto k^{-3}$, 
due to the effect of the Ekman drag \citep{BM17, VBDMC}.
As a reference, the straight grey solid line in figure~{\ref{fig3}} shows the slope of $k^{-4}$ with these axes.
The solid lines in figure~{\ref{fig3}} represent the temporal evolution in the energy spectrum $\Delta E(k,t)$ of
the difference of the fields, ${\bm u}(t) - \tilde{\bm u}(t)$,
where $\tilde{\bm u}(t)$ denotes the velocity field obtained from the CDA with $k_a=4$.
The spectrum $\Delta E(k,t)$ represents the distribution of the synchronisation error, in terms of both amplitude and phase, across the scales.
With increasing thickness, the lines correspond to the results of $\Delta E(k, t)$ at times $ t = 0,\, 10,\, 50,\, 100,\, 150,\, \text{and}\ 200$.
This case is the same as that shown in figure~\ref{fig2}, where $\tilde{\bm q}_0 = 0$ at $t = 0$.
Therefore, $\tilde{\bm q}(t) \simeq 0$ in the initial period of the CDA process, and $\Delta E(k,t) \simeq \langle E(k) \rangle_T~(k/ k_a \ge 1)$.
By definition, 
$\Delta E(k,t) \equiv 0$
for all $k/k_f = k/k_a < 1$ and $t \ge 0$,
since we assume that the `ground-truth' DNS data are available for this spectral range and hence no error exists there.
As expected from the results shown in figure~{\ref{fig2}-\ref{palinstrophy}}, $\Delta E(k,t)$ decreases with time for fixed $k$.

Interestingly, while decaying in magnitude, the shape of the energy spectrum appears to be preserved.
Taking into account the exponential decay characterised by the conditional Lyapunov exponent $\lambda_c$, we assume that the energy spectrum evolves in time as
$\Delta E(k,\,t) \propto \mathcal{E}(k)e^{-2\lambda_c t}$ with some time-invariant function $\mathcal{E}$.
In fact, the scaled energy spectra, $\Delta E(k,\,t)e^{2 \lambda_c t}$, do not change their shape for $t \ge 100$ as shown in the inset,
which suggests the existence of such a time-invariant function $\mathcal{E}(k)$.
Introducing observational data reduces the error, $\Delta E(k,\, t)$ for $k\ge k_a$
almost uniformly across all scales, which may be related to the non-locality of inter-scale interactions discussed later.

\begin{figure}
  \centerline{\includegraphics[width=0.9\linewidth]{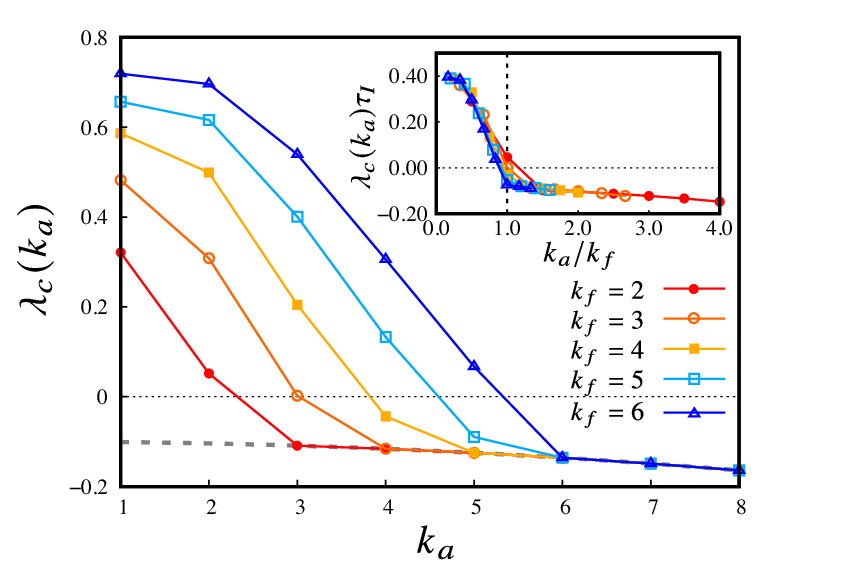}}
  \caption{The conditional Lyapunov exponent $\lambda_c(k_a)$ as a function of $k_a$, for the fixed parameters $\nu = 1.0 \times 10^{-3}$ and $\alpha =1.0 \times 10^{-1}$, and for different forcing wavenumbers $k_f = 2, 3, \dots, 6$, represented by filled red circles, open orange circles, yellow filled squares, open light-blue squares, and blue triangles, respectively.
  The gray dashed curve represents $- \nu k_a^2 - \alpha$.
   The inset shows the same exponents $\lambda_c(k_a)$ normalised by time scale associated with the enstrophy injection $\tau_I$ and the forcing wavenumber $k_f$.}
\label{CLE1}
\end{figure}

Heretofore, we have fixed the forcing wavenumber at $k_f = 4$, with the parameters,
viscosity $\nu =1.0 \times 10^{-3}$, and linear drag coefficient $\alpha = 1.0 \times 10^{-1}$.
As a final test, we investigate the dependence of the synchronisation properties on the particular choice of $k_f, \nu$ and $\alpha$.
Figure~\ref{CLE1} shows the conditional Lyapunov exponent
as a function of $k_a$ with different forcing wavenumbers $k_f=2,\, 3, \, \cdots, \, 6$.
We exclude the case of $k_f=1$, for which chaotic dynamics are not observed.
The conditional Lyapunov exponent for the case of $k_a = 1$ coincides with the maximum Lyapunov exponent, as discussed in \S\ref{subsecCLE}, and is positive for all values of $k_f$, i.e. $\lambda_c(k_a=1) = \lambda_1 > 0$.
With increasing $k_a$, the observational data reduce uncertainty, so that the conditional Lyapunov exponents decrease monotonically and eventually change sign around each $k_f$, i.e. $k_a^{\ast(2D)} = O(k_f)$.

For larger values of $k_a$, the asymptotic behaviour of the perturbation $\delta {\bm q}(t)$ is governed essentially by the viscous and frictional terms,
${\partial}_t \delta {\bm q} \simeq - \nu \Delta \delta {\bm q} - \alpha \delta {\bm q}$,
Consequently, the conditional Lyapunov exponents decrease following $\lambda_c (k_a) \simeq - \nu k_a^2 - \alpha$,
which is represented by the dashed curve shown in figure~\ref{CLE1}.
The data of $\lambda_c (k_a)$ for large $k_a$ lie on the curve $- \nu k_a^2 - \alpha$,
as observed in three-dimensional turbulence by \cite{ISKG}.
Note that, while the conditional Lyapunov exponents shown in figure~\ref{CLE1} are calculated from the long-time average as defined by equation~(\ref{defCLE}), the finite-time dynamics of the perturbation $\delta{\bm q}(t)$ are not necessarily governed solely by such viscous and frictional effects. Further details of the perturbation dynamics will be reported elsewhere.

\begin{figure}
  \centerline{\includegraphics[width=0.9\linewidth]{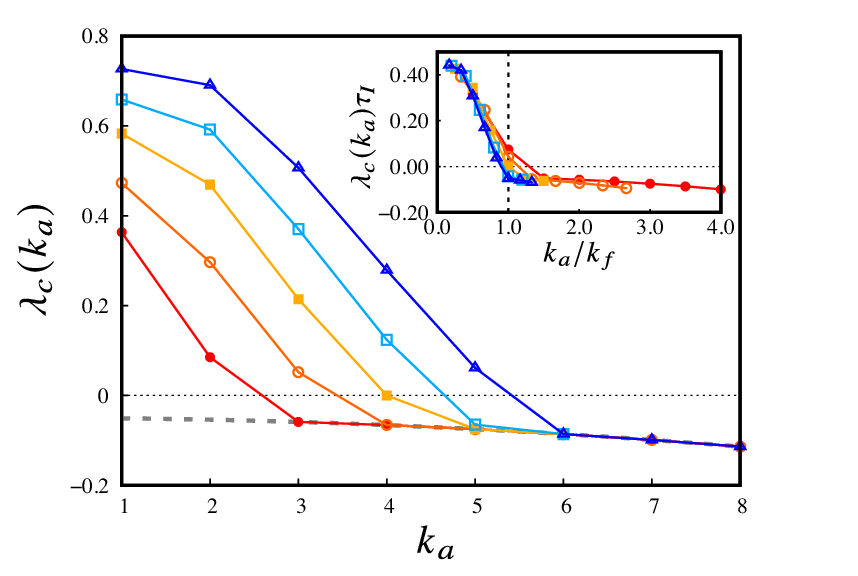}}
  \caption{Same as figure~\ref{CLE1}, for the fixed parameters $(\nu, \alpha) = (1.0 \times 10^{-3},\,5.0\times10^{-2})$.}
\label{CLE2}
\end{figure}

\begin{figure}
  \centerline{\includegraphics[width=0.9\linewidth]{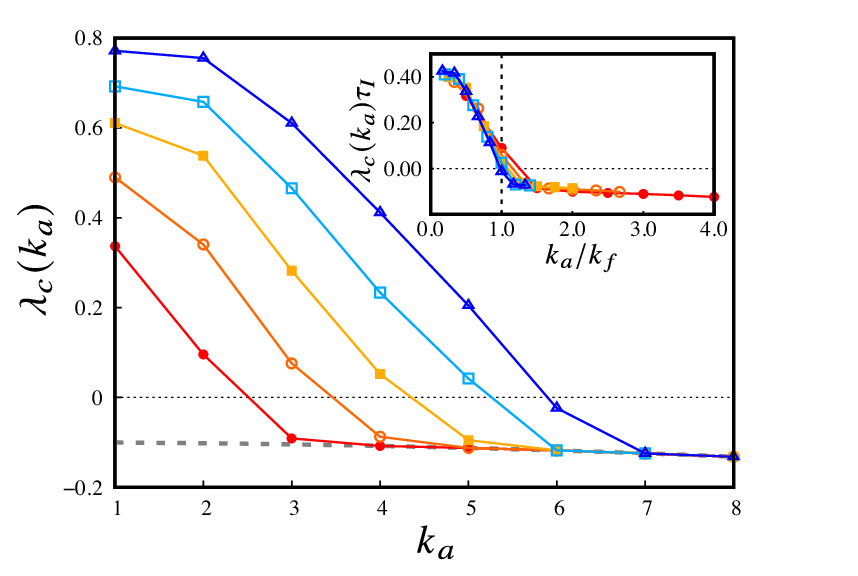}}
  \caption{Same as figure~\ref{CLE1}, for the fixed parameters $(\nu, \alpha) = (5.0\times 10^{-4},\, 1.0 \times 10^{-1})$.}
\label{CLE3}
\end{figure}

The inset of figure~\ref{CLE1} shows the same conditional Lyapunov
exponents normalised by the timescale associated with the enstrophy
injection, $\tau_I$, and the forcing wavenumber $k_f$.
Following \citet{VBDMC}, we define $\tau_I = \eta_I^{-1/3}$, where $\eta_I$ denotes the enstrophy injection rate, given by $\eta_I := - k_f \langle \overline{\omega \cos (k_f y)}\rangle_T$ for the Kolmogorov-type forcing.
Here, the overline denotes the spatial mean.
For statistically steady turbulence, the enstrophy injection rate is balanced by the total enstrophy dissipation rate,
i.e., $\eta_I = \eta_\nu + \eta_\alpha$~\citep{BE}, where $\eta_\nu = 2\nu \langle P \rangle_T$ is the dissipation due to viscosity, and $\eta_\alpha = 2\alpha \langle \Omega \rangle_T$ is that due to Ekman friction.
\cite{VBDMC} identified $\tau_I$ as a key characteristic timescale of two-dimensional Ekman--Navier--Stokes turbulence.
The data shown in the inset of figure~\ref{CLE1} almost collapse onto a single curve, suggesting the scaling
with $\tau_I$ and $k_f$ is appropriate.
Figures~\ref{CLE2} and~\ref{CLE3} show the results of the same
computation as figure~\ref{CLE1}, but with different parameter sets,
$(\nu, \alpha) = (1.0 \times 10^{-3},\,5.0\times10^{-2})$ and $(\nu,
\alpha) = (5.0\times 10^{-4},\, 1.0 \times 10^{-1})$,
respectively. The overall behaviour remains largely unchanged.

The insets of figure~\ref{CLE1}-\ref{CLE3} indicate that $\lambda_c(k_a=1)\tau_I = \lambda_1 \tau_I \simeq 0.40$ for the maximum Lyapunov exponents.
This value is close to the $0.42 \pm 0.01$ which is the maximum Lyapunov exponents normalised by the enstrophy dissipation rate
reported by \citet{Clark} (see their figure~3).
Although their system differs from the present one, for example because they employed hypoviscosity instead of Ekman drag, this similarity may still be worth noting.

\section{Discussion: Dimensional dependence of synchronisation properties}
\label{sec:dimensional-dependence}
The results presented in the previous section strongly suggest that in two-dimensional turbulence as examined here,
the critical wavenumber, the lower bound on the resolution of observational data required for data assimilation, coincides with the forcing scale, i.e. 
$k_a^{\ast(2D)} = {O}(k_f)$.
In other words, in two-dimensional turbulence with a forward
  enstrophy cascade,
  the number of degrees of freedom associated with the large-scale structures governing the whole dynamics
is substantially smaller than the total degrees of freedom.

Notably, the critical wavenumber $k_a^{\ast(2D)}$ is located at the {\it upper edge} of the power-law scaling range.
Interestingly, in three-dimensional turbulence in a periodic box, the equivalent critical wavenumber $k_a^{\ast(3D)} \simeq 0.2 \eta^{-1}$, differs markedly from that in the two-dimensional turbulence discussed above.
The critical wavenumber $k_a^{\ast(3D)}$ actually lies near the {\it lower edge} of the inertial range, i.e., close to the Kolmogorov length scale,
and clearly exhibits a dependence on the viscosity $\nu$.
As turbulence develops further, achieving synchronisation necessitates
increasingly well-resolved observational data in 3D, but not
  apparently in 2D.

To account for the observed dimensional differences in this synchronisation property of NS turbulence, we propose the following physical interpretation.
We begin by interpreting the fact that $k_a^{\ast(3D)} \simeq 0.2\eta^{-1}$ in three-dimensional turbulence.
Specifically, we consider why synchronisation fails when $k_a \ll k_a^{\ast(3D)}$;
for instance, in a setting where the observational data are restricted to the energy-containing range.
In three-dimensional turbulence, interscale interactions are known to be {\it scale-local}.
Turbulent structures at a given scale predominantly generate smaller-scale structures, specifically, those at approximately $0.58$ times the original scale \citep{YGT},
which then successively generate smaller and smaller scale structures.
Turbulence flow structures are thus generated (at least on average) sequentially from large to small scales, the physical manifestation of the (forward) energy cascade of three-dimensional turbulence.
Taking this into consideration, it may be expected that continuously providing observational data at the largest scales would induce the successive formation of smaller-scale structures.
The information contained in the observational data is `transmitted'
progressively to smaller scales.

Crucially however, that transmission inevitably involves amplification
of uncertainty when $k_{\mathrm{a}} \ll k_a^{\ast(3D)}$, as data
assimilation fails to reconstruct small-scale structures
quantitatively accurately. A key mechanism underlying this failure is the inherent \textit{orbital instability} of three-dimensional turbulence.
As noted in the Introduction, the maximum Lyapunov exponent of three-dimensional turbulence corresponds to the inverse of the Kolmogorov timescale, capturing the physical phenomenon that the fastest time scales in turbulence inevitably introduce uncertainty. 
The most unstable spatial mode that amplifies small errors, the Lyapunov vector associated with the maximum Lyapunov exponent, $\lambda_1$,
exhibits a peak in Fourier space
at $k \simeq 0.2\eta^{-1}$ in the energy spectrum~\citep{BM17,BK}.
Furthermore,
a study based on the GOY shell model \citep{YO88} suggests that
the peaks of the less unstable modes, associated with
the subsequent positive Lyapunov exponents, $\lambda_i>0~(i \ge 2$),
occur at wavenumbers lower than that of the most unstable mode.
Hence, all unstable modes
associated with the positive Lyapunov exponents exist only at wavenumbers {\it below} $0.2\eta^{-1}$.
Combining this with the previously described picture of the energy cascade,
when $k_{\mathrm{a}} \ll k_a^{\ast(3D)}$,
while the observational information of large scales is transmitted towards smaller scales
in the DA process, the unstable spatial modes amplify small-scale errors exponentially. 
The `inverse cascade' of errors~\citep{BM17} toward larger scales inevitably disrupts the transmitted information,
and therefore, the data assimilation fails when $k_{\mathrm{a}} < k_a^{\ast(3D)}$.
On the other hand, when the observational data include all unstable modes, i.e., $k_{\mathrm{a}} > k_a^{\ast(3D)}$,
there now exists no mechanism by which errors can be amplified,
and so data assimilation is successful.

In the case of two-dimensional turbulence, the physical mechanism underlying the critical wavenumber $k_a^{\ast(2D)} = O(k_f)$
is qualitatively different, although perhaps not as clear as in  the three-dimensional case.
The properties of two-dimensional turbulence can strongly depend on the details of the forcing and the energy dissipation mechanism~\citep{BE},
and few studies have examined its orbital instability.
Consequently, understanding the origin of the critical wavenumber through this particular lens is challenging.
Nevertheless, it is reasonable to expect that the {\it nonlocality} of inter-scale interactions \citep{Okitani} and the orbital instability play key roles in two-dimensional flows. 
Nonlocal inter-scale interactions dominate in two-dimensional turbulence, and indeed it is well-known that there is an {\it inverse} energy cascade in two-dimensional turbulence \citep{BE}, at least suggestive of the fact that larger scale structures `know' about smaller scale structures that are `pumping' energy up-scale. 
As a result, even when observational data are available only in the energy-containing range (i.e., $k_{{a}} = O(k_{f})$),
information can be transmitted across scales via nonlocal
interactions, making it possible to reconstruct directly small-scale
structures.
It appears, at least heuristically, that the required
small scales structures to cascade energy upscale to form (and maintain) the
observed larger scale structures emerge (and survive) spontaneously in
a way that remains consistent with the downscale enstrophy cascade.  
Furthermore, \cite{YO88} suggests that
all unstable modes
associated with the positive Lyapunov exponents in two-dimensional turbulence
exist only at wavenumbers {\it below} $O(k_{f})$.
Therefore, when the observational data include all unstable modes, i.e., $k_{\mathrm{a}} \ge O(k_{f})$,
there exists no mechanism for error amplification,
and small-scale structures are directly reconstructed via the nonlocal
inter-scale interactions (again consistently with the inverse energy
cascade and the forward enstrophy cascade);
that is, data assimilation can be successful.

\section{Conclusion}\label{Conclusion}
Synchronisation in turbulence is a key property to understand turbulence dynamics,
and has been studied extensively for three-dimensional turbulence.
The present study initiates a new direction of research into {\it two}-dimensional turbulence by introducing a novel approach based on synchronisation.
Specifically, through the use of the data assimilation method and Lyapunov analysis, we have demonstrated that the essential resolution of  observations for flow field reconstruction in forced two-dimensional turbulence is surprisingly lower than the equivalent essential resolution in forced three-dimensional turbulence.
More precisely, in the two-dimensional turbulent flows studied here, the critical wavenumber for reconstruction is close to the forcing scale, i.e., $k_{a}^{\ast(2D)} = {O}(k_{f})$, while the critical wavenumber for the equivalent three-dimensional turbulent flows
 is close to the dissipation scale, i.e., $k_{a}^{\ast(3D)} \simeq 0.2/\eta$.
We further propose  physical mechanisms underlying the difference in synchronisation properties, based on
the (non-)locality of interscale interactions and orbital instabilities in turbulence dynamics.

Compared with three-dimensional turbulence, the understanding of synchronisation properties from the perspective of turbulence physics
remains somewhat more speculative in two-dimensional flows, despite their importance.
Also, while we have identified the critical wavenumber $k_{a}^{\ast(2D)}$ for two-dimensional Kolmogorov flows with Ekman drag, establishing the generality of this result
through extensions of the present work remains an important challenge for future research.
Only recently have the fundamental properties of Ekman--Navier--Stokes turbulence, namely the correction to the energy spectrum, been clarified using GPU-based computations over a wide range of parameters of $\nu$ and $\alpha$~\citep{VBDMC}.
Although the calculation of the conditional Lyapunov exponents $\lambda_c$ requires a considerable additional computational cost beyond that of the DNS, extending the present analysis to a broader parameter range will be an important next step.
In particular, clarifying the relationship between the timescale associated with synchronisation, characterised by $\lambda_c$, and the relevant timescales within turbulence will provide valuable insight into the fundamental understanding of turbulence.

Beyond this, it is crucial to study further the synchronisation properties of two-dimensional turbulence with, for example, a more detailed consideration of the inverse energy cascade and realistic boundary conditions. 
In particular, studying synchronisation in two-dimensional turbulence that exhibits a {\it double} cascade~\citep{BE} is intriguing.
In three-dimensional turbulence, \citet{ClarkDiLeoni2020} identified the critical wavenumber $k_a^{\ast(3D)} \simeq 0.2/\eta$ as the characteristic scale where the nonlinear and viscous energy fluxes are in balance.
A complementary approach combining such flux-based and Lyapunov analyses is crucial for elucidating the dynamics of two-dimensional turbulence.
Furthermore, it would be particularly interesting to consider flows with other physical mechanisms such as rotation and stratification that can lead, at least in some regimes, to quasi-two-dimensional dynamics, providing a natural test-bed to investigate transitions from behaviour associated with the two `pure' end members of exactly two-dimensional and inherently three-dimensional flows. 
In this context, it will be valuable to summarise
  systematically the critical wavenumbers associated with synchronisation during such transitions from two- to three-dimensional systems, using the framework developed in this study.
Finally, it is also essential to develop a quantitative framework that solidifies the interpretation presented in the previous section, where quantifying information transmission across scales in turbulent dynamics,
as explored by, for example \cite{TA,TA25},
will play a crucial role.
Pursuing this direction could offer valuable insight into the fundamental nature of turbulence.

\backsection[Supplementary material]{\label{SupMat}A supplementary video is available at \\https://doi.org/10.1017/jfm.2019...
The upper part of the video corresponds to the vorticity fields shown in figure~{\ref{fig2}}(a) (from left to right: the velocity field obtained from DNS, $\bm{u}(t) = \bm{p}(t) + \bm{q}(t)$; the observational data $\bm{p}(t)$; and the data assimilation result $\tilde{\bm{u}}(t) = \bm{p}(t) + \tilde{\bm{q}}(t)$). The lower part displays the corresponding enstrophy. In particular, the blue solid line indicates the enstrophy $\Omega(t)$ of the DNS solution, the blue dashed line represents the approximate enstrophy arising from the simulation implementing the CDA process, and the red solid line shows the enstrophy error denoted $\Delta\Omega(t)$ and defined in (\ref{eq:errordef}).}

\backsection[Acknowledgements]{
The authors are grateful to the three anonymous reviewers whose detailed and constructive comments helped improve the focus and overall quality of the manuscript.
M. I. gratefully acknowledges M. Yamada and E. S. Titi for their
insightful discussions, and the hospitality of the Department of
Applied Mathematics and Theoretical Physics at the Centre for Mathematical Sciences, University of Cambridge, where this work was carried out during his sabbatical leave.
Part of the direct numerical simulations of the Navier--Stokes equations was conducted using
the supercomputer systems at the Japan Aerospace Exploration Agency (JAXA-JSS2).}

\backsection[Funding]{
This work was partially supported by JSPS Grants-in-Aid for Scientific Research (Grants 
No.~24H00186,~No.~22K03420,~and No.~22H05198). }

\backsection[Declaration of interests]{The authors report no conflict of interest.}

\backsection[Data availability statement]{The data that support the findings of this study are available from the corresponding author upon reasonable request.}

\backsection[Author ORCIDs]{
M. Inubushi, https://orcid.org/0000-0002-7541-8458;
C. P. Caulfield, https://orcid.org/0000-0002-3170-9480
}


\appendix


\bibliographystyle{jfm}


\end{document}